\documentclass{article}
\usepackage{fortschritte}
\usepackage{amsfonts}
\def\beq{\begin{equation}}                     %
\def\eeq{\end{equation}}                       %
\def\bea{\begin{eqnarray}}                     
\def\eea{\end{eqnarray}}                       
\def\NP{{\it Nucl. Phys.} }                    
\def\PL{{\it Phys. Lett.} }                    
\def\ap{\alpha ^{\prime}}                      %
\begin {document}
\def\titleline{
%
%
%
Effective Actions Near Singularities:
The STU-Model \footnote{Contribution 
to the proceedings of the 35th Symposium Ahrenshoop, August 2002.
Based on \cite{MZ,LMZ}.} 
}
\def\email_speaker{
{\tt
%
%
Marco.Zagermann@cern.ch             
}}
\def\authors{
%
%
%
%
%
Marco Zagermann\sp                           
}
\def\addresses{
%
%
%
%
                             %
CERN, Theory Division\\            %
CH-1211 Geneva 23, Switzerland\\            
}
\def\abstracttext{
%
%
We derive the low energy effective action of the $STU$-model in
four and five dimensions near the line $T=U$, where $SU(2)$ gauge
symmetry enhancement occurs. By `integrating in' the light $W^{\pm}$
bosons together with their superpartners, the quantum corrected
effective action becomes non-singular at $T=U$ and manifestly
$SU(2)$ invariant. The  four-dimensional theory  is found to be
consistent with modular invariance and the five-dimensional
decompactification limit.

}
\large
\makefront
\section{Introduction}
One of the, at first sight,    somewhat irritating  
features of string theories (or M-theory) is
that their mathematical consistency requires the existence
of more than four  space-time dimensions. This apparent discrepancy
with observation is conventionally accomodated by demanding the
additional dimensions to be compact. Assuming the compact part of space-time
and the string length
to be sufficiently small, the resulting spectrum allows a
clear separation between very heavy modes and modes that are very
light, or massless.
 If one is only interested in low energy processes, the heavy
modes can be integrated out, and one is left with an effective,
lower-dimensional field theory that describes all low-energy
phenomena.

The set of light fields  of such a string compactification
typically includes a number of scalar fields whose vacuum
expectation values (vevs) are  not fixed by a perturbative scalar
potential. The vevs of some of these moduli fields have a simple
geometrical interpretation in that they parameterize deformations
of the compactification manifold that cost no energy.

For certain values of these deformation parameters, one sometimes
encounters the phenomenon that some of the generically heavy
modes suddenly become light. As a classical example, consider  a  string
compactification  that involves an $S^1$ factor. The radius,
$R$, of the circle is in general not fixed by perturbative string
physics and thus gives rise to a modulus, $\phi$, whose vev is
proportional to the radius:
 $ \langle \phi \rangle = aR$. At $\langle \frac{\phi}{a} \rangle 
= \sqrt{\ap}$,
the circle is at its self-dual radius, and certain otherwise
massive Kaluza-Klein and winding modes become massless (and lead
to an $SU(2)$ gauge symmetry enhancement).

As another,  non-perturbative,
  example (which can, however, sometimes be
dual to the previous one  (see below)), consider a
compactification manifold with a non-trivial $p$-cycle. If the
string (or M-)theory in question has appropriate solitonic objects
($p$-branes), 
these may wrap around the cycle and give rise
to pointlike particles in the uncompactified dimensions
\cite{Strom}. The mass of such  particles is 
proportional to the
volume of the $p$-cycle. When the size of the cycle corresponds to
a modulus in the low energy effective theory, a vanishing vev for
this modulus field corresponds to the cycle being collapsed to
zero size. The wrapped brane then gives rise to additional
massless states in the lower-dimensional theory at that particular
point in the moduli space.

Clearly, away from these special points, such extra states are
heavy and should be integrated out of the effective action. Near
the points in moduli space where they become light, however, it is
inaccurate, or rather  inconsistent, to neglect them in the low
energy theory. This inconsistency is typically
 reflected in couplings becoming singular or discontinuous when
the modulus reaches the special value at which additional fields
become massless. In order to obtain a consistent and non-singular
action, one would have to avoid integrating out the extra light
modes, at least in the region of the moduli space where they are
light.

This can, in principle, be achieved in two different ways:\\
(i) Either one uses a microscopic string  calculation in order to
determine all the low energy couplings of the extra modes, or\\
(ii) one hopes that the generic low energy effective action
without the extra states is sufficiently well-known and that there
are sufficiently many symmetries involved so that one can
``integrate the extra modes back in'' by simply exploiting all
these symmetries.

The second (``bottom-up'') method might not always be applicable,
but if it is, it can be much simpler than the first, especially,
when the microscopic formulation is not so well understood (such
as, e.g., in the case of M-theory). In this talk I summarize two
non-trivial examples \cite{MZ,LMZ} in which the second method has
been carried out in full detail. For applications in the context
of geometric phase transitions, I refer to T. Mohaupt's talk
\cite{Moh}.

\section{The $STU$-model in  five dimensions}
The model we are going to study first is the $E_8\times E_8$
heterotic string on $K3\times S^1$ with instanton numbers
$(14,10)$ \cite{AFT}. This model is believed to be dual
\cite{PT,CCDF} to M-theory compactified on the Calabi-Yau
threefold $Y_{1,1,2,8,12}(24)$ \cite{KLM,HKTY,KV}, which is an
elliptic fibration over the second Hirzebruch surface
$\mathbf{I\! F}_2$. The generic low energy effective action of
this compactification describes the coupling of two Abelian vector
multiplets and 244 neutral hypermultiplets to five-dimensional
(5D), $\mathcal{N}=2$ (i.e., minimal) supergravity. The
hypermultiplets play no r\^{o}le in the following and will be
consistently truncated out. The bosonic part of the Lagrangian is
of the form
\begin{eqnarray}
\label{Lagrange1} e^{-1}\mathcal{L}_{\rm bosonic}&=&-\frac{1}{2}R-
\frac{1}{4} {\stackrel{\circ}{a}}_{IJ} F_{\mu\nu}^{I}F^{\mu\nu J}-
\frac{1}{2} g_{xy}(\partial_{\mu}\phi^{x})(\partial^{\mu}
\phi^{y})\nonumber\\
&& +\frac{e^{-1}}{6\sqrt{6}}C_{IJK}
\varepsilon^{\mu\nu\rho\sigma\lambda}F_{\mu\nu}^{I}
F_{\rho\sigma}^{J}A_{\lambda}^{K},
\end{eqnarray}
where $\phi^x$ $(x=1,2)$ are two real scalars, and the index
$I=0,1,2$ collectively labels the graviphoton and the vector
fields from the two vector multiplets.
 The completely symmetric tensor $C_{IJK}$ in the $FFA$ term
of (\ref{Lagrange1}) is independent of the scalar fields and
completely determines the entire theory  via a cubic polynomial
(or ``prepotential'') \cite{GST1}
\begin{equation}
\mathcal{V}(h):=C_{IJK}h^{I}h^{J}h^{K}
\end{equation}
in three real variables $h^{I}$ ($I=0,1,2$). $\mathcal{V}(h)$
endows the auxiliary space ${\mathbb{R}}^{3}$ spanned by the $h^I$
with a metric,
\begin{equation}\label{aij}
a_{IJ}(h):=-\frac{1}{3}\frac{\partial}{\partial h^{I}}
\frac{\partial}{\partial h^{J}} \ln \mathcal{V}(h).
\end{equation}
The  two-dimensional    target space, $\mathcal{M}$, of the scalar
fields $\phi^x$ can then be represented as the hypersurface
\cite{GST1}
\begin{equation}\label{hyper}
{\cal V} (h)=C_{IJK}h^{I}h^{J}h^{K}=1
\end{equation}
with $g_{xy}$ being the pull-back of (\ref{aij}) to $\mathcal{M}$.
The quantity ${\stackrel{\circ}{a}}_{IJ}(\phi)$ appearing in
(\ref{Lagrange1}), finally, is given by the restriction of
$a_{IJ}$ to $\mathcal{M}$: $
{\stackrel{\circ}{a}}_{IJ}(\phi)=a_{IJ}|_{{\cal V}=1} $. Due to
these relations, the physical requirement that
 $g_{xy}$ and ${\stackrel{\circ}{a}}_{IJ}$ be positive definite
 imposes  constraints
on the possible prepotentials $\mathcal{V}$.

For our particular string compactification, these constraints are,
of course, satisfied, and the corresponding prepotential reads
\cite{AFT} \beq \label{STU1}
\mathcal{V}(S,T,U)=STU+\frac{1}{3}U^3, \eeq where, as is common in
the literature, the letters $S,T,U$ are used instead of the
variables $h^0,h^1,h^2$. This prepotential is valid in the region
$T>U$. For $T<U$, it has to be replaced by \cite{AFT,MZ} \beq\label{STU2}
\mathcal{V}(S,T,U)=STU+\frac{1}{3}U^3 +\frac{1}{3}(T-U)^3.\eeq
Obviously, (\ref{STU1}) and (\ref{STU2}) combine
to form a continuous function $\mathcal{V}$  at
$T=U$. The couplings in the Lagrangian (\ref{Lagrange1}), however,
depend on derivatives of $\mathcal{V}$, and these are not all
continuous at $T=U$. The physical reason for these discontinuities is
exactly as  described in the Introduction: At $\langle T-U \rangle
=0$, the circle of the heterotic compactification manifold
$K3\times S^1$ is at its self-dual radius, and two additional
vector multiplets (containing two $W^{\pm}$ bosons) become light
and restore an $SU(2)$ gauge symmetry. In the dual M-theory
picture, this very same situation corresponds to a collapsed
2-cycle in the Calabi-Yau threefold, and the two additional vector
multiplets are supplied by the zero modes of wrapped M2 branes
\cite{EW}. Thus, near $T=U$, a complete low energy effective
theory should also contain these two additional vector multiplets.
Let us use $\mathcal{L}_{\rm in}$ do denote the Lagrangian of this
more complete theory (``in'', because the two extra vector
multiplets have been ``integrated back in''). $\mathcal{L}_{\rm
in}$ has to
have the following properties:\\
 (i) Just as $\mathcal{L}$ in eq.
(\ref{Lagrange1}), it is based on a cubic polynomial, but now this
is a polynomial in $3+2=5$ variables. We call this polynomial
$\mathcal{V}_{\rm in}$.\\
(ii) Whereas $\mathcal{L}$ is Abelian, $\mathcal{L}_{\rm in}$
exhibits $SU(2)$ as a Yang-Mills-type gauge symmetry. This implies
that three of the five variables entering $\mathcal{V}_{\rm in}$
transform in the adjoint of $SU(2)$; the other two are $SU(2)$
singlets. We use $C^a$ $(a=1,2,3)$ to denote the $SU(2)$ 
triplet and $Z^1$
and $Z^2$ for the two singlets. The polynomial
$\mathcal{V}_{\rm in}(Z^1,Z^2,C^a)$ has to be $SU(2)$
invariant. \\
(iii) The metrics $g_{xy}$ and $ {\stackrel{\circ}{a}}_{IJ}$ that
follow from $\mathcal{V}_{\rm in}$ have to be positive definite.\\
(iv) Integrating out the two extra multiplets from
$\mathcal{L}_{\rm in}$ should reproduce $\mathcal{L}$ and the
underlying prepotential $\mathcal{V}(S,T,U)$ (eqs. (\ref{STU1}),
(\ref{STU2})).

In \cite{MZ}, it was shown that, modulo reparameterizations,
$\mathcal{V}_{\rm in}(Z^1,Z^2,C^a)$ (and with it  $\mathcal{L}_{\rm
in}$) is completely fixed by (i)-(iv). The approach was to view
the integrating out process $\mathcal{V}_{\rm in}
 \rightarrow \mathcal{V}$  as a two-step procedure, in which, in
 the language of Feynman diagrams, the two extra multiplets
 represented by, say,  $C^1$ and  $C^2$ are first removed as external
 lines by setting them equal to zero in $\mathcal{V}_{\rm in}$.
 This yields an intermediate (unphysical) prepotential
\beq \mathcal{V}_{\rm in}^{\rm truncated} :=\mathcal{V}_{\rm
in}|_{C^1=C^2=0}.\eeq
In a second step, one then has to take into
account that the two extra multiplets can also occur as internal
lines and run in loops. Integrating them out will thus also
produce additional effective interactions among the remaining
fields. These induced interactions are subsumed in an additional
contribution $\delta \mathcal{V}$, so that \beq
\mathcal{V}=\mathcal{V}_{\rm in}^{\rm truncated}+\delta \mathcal{V}.
\eeq In order to determine $\mathcal{V}_{\rm in}$, one simply has
to go backwards. As explained in \cite{EW,MS},  $\delta
\mathcal{V}$ is simply given by a one-loop correction to the $FFA$
term in $\mathcal{L}$, which in our case turns out to be $\delta
\mathcal{V}=-(T-U)^3/6$, implying \beq \mathcal{V}_{\rm in}^{\rm
truncated}= STU +\frac{1}{3}U^{3}+\frac{1}{6}(T-U)^3.\eeq This
truncated prepotential now has to be ``untruncated'', i.e., the
variables $(S,T,U)$ have to be transformed to $(Z^1,Z^2,C^3)$, and
$C^{1}$ and $C^{2}$ have to be appropriately re-inserted. Using the
list of admissible polynomials given in \cite{EGZ}, one can show
that, modulo obvious linear transformations, there is essentially 
only one way to do that \cite{MZ}: The two $SU(2)$ singlets are given
by $Z^1=S-(T-U)/2$ and $Z^2=(T+U)/2$, whereas $C^3=(T-U)/2$. In
terms of these variables, $\mathcal{V}_{\rm in}^{\rm truncated}$
becomes quadratic in $C^{3}$, and re-introducing $C^1$ and $C^2$
is simply done by $SU(2)$ covariantization: $(C^3)^2 \rightarrow
[(C^1)^2+ (C^2)^2+(C^3)^2]$.

\section{The  $STU$-model in four dimensions  }
In the previous section, we have reconstructed the low energy
effective theory of a five-dimensional string (or M-theory)
compactification near an $SU(2)$ enhancement line in the moduli
space. The one-loop threshold effects allowed by $\mathcal{N}=2$
supersymmetry added a certain degree of non-triviality to this
exercise. Nevertheless, one might wonder how much this
construction relied on the purely cubic form of the prepotential
in five dimensions. Let us therefore, in this section, consider
the same theory compactified to four dimensions. In other words,
we are now considering the heterotic string on $K3\times T^2$ with
instanton numbers (14,10) or, equivalently, type IIA string theory
on the Calabi-Yau manifold $Y_{1,1,2,8,12}(24)$
\cite{KLM,KV,DKLL,AFGNT,HM}. The generic low energy effective  theory
describes 4D, $\mathcal{N}=2$ supergravity coupled to three
Abelian vector multiplets and 244 neutral hypermultiplets. Again,
the hypermultiplets can be ignored, and the vector multiplet
couplings are  summarized in terms of a prepotential,
$\mathcal{F}$ \cite{dWvP}. Just as in five dimensions, this
prepotential depends on three variables $(S,T,U)$. This time,
however, the fields $(S,T,U)$ are complex, rather than real, and
they do not have to satisfy a hypersurface constraint of the form
(\ref{hyper}). Furthermore, the prepotential $\mathcal{F}$ no
longer has to be cubic. Instead, $\mathcal{F}$ can now be a rather
arbitrary holomorphic function of the moduli (with possible
logarithmic branch cuts and singularities). For $\textrm{Re
}S>\textrm{Re }T>\textrm{Re }U>0$, $\mathcal{F}$ turns out to be
\cite{HM} \beq\label{neueVersion} {\cal F} = STU + \frac{1}{3} U^3
+ \frac{2}{(2\pi)^3} Li_3 \left( e^{-2 \pi (T-U)} \right) +
\frac{2}{(2\pi)^3} \sum_{k,l=0}^{\infty} c_1(kl) Li_3 \left( e^{-2
\pi (kT + lU)} \right) +\mathcal{F}^{(NP)} \;, \eeq where $Li_3$
denotes the third polylog (see, e.g., \cite{HM}), and the
coefficients $c_1(kl)$ can be found in \cite{HM}. The term
$\mathcal{F}^{(NP)}$ summarizes  non-perturbative corrections.

 The prepotential (\ref{neueVersion}) has   a logarithmic 
singularity along the
 surface
$T=U$, where two additional vector multiplets become massless and
enhance the gauge group to $U(1)^3\times SU(2)$. At $T=U=1$ and
$T=U=e^{i\pi/6}$, the gauge group is further enhanced to,
respectively, $U(1)^2\times SU(2)^2$ and $U(1)^2\times SU(3)$. In
contrast to the 5D case,  these symmetry enhancements do not
survive non-perturbative quantum corrections \cite{KLM,SW}, and we
therefore have to restrict our considerations to the perturbative
heterotic string, dropping from now on the term
$\mathcal{F}^{(NP)}$.

We will now try to derive an effective action, $\mathcal{L}_{\rm
in}$, that includes the two additional light vector multiplets
near $T=U$ and describes the $SU(2)$ gauge symmetry enhancement
\cite{LMZ}. Except at the discrete points where further gauge
symmetry enhancement occurs, we expect this effective theory to be
non-singular. The underlying prepotential will be called
$\mathcal{F}_{\rm in}$. Just as its  five-dimensional analogue,
$\mathcal{F}_{\rm in}$ is a function of 3+2=5 variables: one
$SU(2)$ triplet $C^{a}$ $(a=1,2,3)$ and two singlets, $Z^1$ and
$Z^2$. Again, we expect $\mathcal{F}$ and $\mathcal{F}_{\rm in}$
to be related by a relation of the form \beq
\label{Fgleichung}\mathcal{F}=\mathcal{F}_{\rm
in}^{\rm truncated}+\delta \mathcal{F}, \eeq 
where $ \mathcal{F}_{\rm in}^{\rm
truncated}:=\mathcal{F}_{\rm in}|_{C^1=C^2=0}$, and $\delta
\mathcal{F}$ subsumes the threshold effects.

Let us introduce $T_{\pm}:=(T\pm U)/2$. Obviously, $T_{-}$
represents the order parameter of the symmetry breaking $SU(2)
\rightarrow U(1)$, and when we identify $(C^1\pm i C^2)$ with the
scalar superpartners of the $W^{\pm}$-bosons (as we implicitly did
above), we have to identify $C^3=T_{-}$. The mass of the $W^{\pm}$
bosons  is then proportional to the vev of $|T_{-}|$. Integrating
them out induces a one-loop threshold correction to the gauge
coupling of the vector field $A_{\mu}^{-}$, i.e.,  the
superpartner of the scalar field $T_{-}$. This correction is of
the form \cite{DKLL,KL,Wein} $\delta g^{-2}\sim log |m|^2 \sim log
|T_{-}|^2$. The quantity $g^{-2}$ is essentially the second
derivative of the prepotential, and if one inserts all prefactors
correctly\cite{LMZ}, one deduces that \beq \label{deltaF} \delta
\mathcal{F} = -\frac{2}{\pi}T_{-}^2\log T_{-} +A_{2}T_{-}^2 +
A_{1}(T_{+})T_{-}+A_{0}(T_{+}). \eeq Here, $A_{2}$ is an arbitrary
constant related to the cut-off scale, and 
$A_{1}(T_{+})$ and $A_{0}(T_{+})$ are a priori undermined functions.

Using (\ref{neueVersion}) and (\ref{Fgleichung}), one can now
solve for $\mathcal{F}_{\rm in}^{\rm truncated}$. Expanding the
first polylogarithmic term in (\ref{neueVersion}), one then finds
that the  logarithmic singularity in $\mathcal{F}$ is precisely
cancelled by $\delta \mathcal{F}$ as given in (\ref{deltaF})
\cite{LMZ}. Thus, $\mathcal{F}_{\rm in}^{\rm truncated}$ is
regular at $T=U$, as it should.

It remains to ``untruncate'' $\mathcal{F}_{\rm in}^{\rm
truncated}$ to obtain the desired function $\mathcal{F}_{\rm in}$.
Just as in five dimensions, this is done by first going over from
the variables $(S,T,U)$ (or $(S,T_{+},T_{-})$) to the $SU(2)$
covariant variables $(Z^1,Z^2,C^3)$ and then replacing everywhere
$(C^3)^2$ by $[(C^1)^2+(C^2)^2+(C^3)^2]$. Modulo obvious linear
combinations, the right change of variables turns out to be the
same as in five dimensions \cite{LMZ}: 
$Z^1=S-T_{-}$, $Z^2=T_{+}$, $C^3=T_-$.
Modulo theta angles, one then finds that \cite{LMZ}, in terms of
the variables $(Z^1,Z^2,C^3)$, $C^3$ only appears with \emph{even}
powers, provided that the as yet undetermined function
$A_{1}(T_{+})$ vanishes identically: $A_{1}(T_{+})\equiv 0$. The
substitution $(C^3)^2  \rightarrow  [(C^1)^2+(C^2)^2+(C^3)^2]$ can
then be readily performed.

It remains to determine the remaining unknown function
$A_{0}(T_{+})$  that appeared in  eq. (\ref{deltaF}). As is shown in
\cite{LMZ}, a diagonal $SL(2,\mathbb{Z})$ subgroup of the
perturbative quantum symmetry $SL(2,\mathbb{Z})_T\times
SL(2,\mathbb{Z})_U$ remains unbroken on the line $T=U$  in the
`in' theory and implies $\partial_{+}^5 A_{0}(T_+)\equiv 0$.
Hence, $A_{0}(T_+)$ can at most be a quartic polynomial in $T_+$.
Taking the 5D decompactification limit of $\mathcal{L}_{\rm in}$,
one can then even show that $A_0(T_+)$ can  be at most quadratic
in $T_+$ \cite{LMZ}.

To sum up, up to a polynomial of the form $A_0(T_+)=a_0+a_1 T_+ +
a_2 T_+^2$, the prepotential  $\mathcal{F}_{\rm in}$ can be
reconstructed using only symmetry arguments and some 4D quantum
field theory reasoning. In order to fix the remaining three
coefficients, a few more couplings have to be considered. In any
case, it shows that the method we used successfully in 5D, can
also be applied in 4D.

{\bf Acknowledgement} The results presented in this talk were
obtained in collaboration with Jan Louis and Thomas Mohaupt
\cite{MZ,LMZ}. The work was  supported in part by the German
Science Foundation (DFG).

\end{document}